\documentclass[prl,aps,twocolumn]{revtex4}
\begin{document}
\title{Locking information in black holes}
%\author{John A. Smolin$^1$ and Jonathan Oppenheim$^2$}
\author{John A. Smolin}
\email{smolin@watson.ibm.com}
\affiliation{IBM T.J. Watson Research Center, Yorktown Heights, NY 10598, 
U.S.A.}
\author{Jonathan Oppenheim}
\email{jono@damtp.cam.ac.uk}
\affiliation{Department of Applied Mathematics and Theoretical Physics, 
University of Cambridge, Cambridge CB3 0WA, U.K.}

\begin{abstract}
The black hole information loss paradox has plagued physicists
since Hawking's discovery that black holes
evaporate thermally in contradiction to the unitarity expected by 
quantum mechanics.
%The calculation suggests that
%information thrown into a black hole is evaporated away as thermal
%radiation and is destroyed: either the unitary laws of quantum
%theory break down, or we must modify the laws of general relativity.
Here we show that one of the central presumptions of the debate is
incorrect. Ensuring that information not escape during the
semi-classical evaporation process does {\em not} require that all the
information remain in the black hole until the final stages of
evaporation.  Using recent results in quantum
information theory, we find that the amount of information that must
remain in the black hole until the final stages of evaporation can be
very small, even though the amount 
%of information which has 
already radiated away is negligible.  Quantum effects mean that information
need not be additive: a small number of quanta can {\it lock} a large
amount of information, making it inaccessible.  When this small number
of locking quanta are finally emitted, the full information (and
unitarity) is restored.  Only if the number of initial states
is restricted will the locking mechanism leak out information early.
\end{abstract}
\maketitle

The laws of quantum mechanics and quantum field theory ensure
predictability---if we completely specify the initial system, and
know all the interactions, then we can know the state of the system at
all future times.  All the known laws of physics satisfy this
principle, called unitarity, with one glaring exception: Hawking 
showed \cite{HawkingOriginal} that a black
hole apparently cause this predictability to break down.  If we have a
completely specified system which forms a black hole, and we let the
black hole evaporate, then Hawking's calculation states that it should
evolve into thermal radiation, which is a maximally random and
unpredictable---information is lost\cite{preskill-infoloss-note}.

Many general relativists regard this loss of predictability as
inevitable, contending there is no paradox.  After all, one can make
models where evolution is non-unitary \cite{unruh-wald-nonu}.  On the
other hand, particle physicists, who
cherish unitarity, traditionally insist that the laws of
general relativity cannot be strictly true and that unitarity must be
preserved.  Famously, John Preskill bet Stephen Hawking and Kip Thorne
that information is not lost in black holes.  And while some may have
changed their minds \cite{hawking-gr17}, few would argue the situation
resolved, and the mystery is, if anything, more pronounced.  If
unitarity is preserved, how is it preserved, and if it is lost, how
are the laws of quantum mechanics modified?

The essential problem is that quantum mechanics is a unitary theory
and the idea of an essentially irreversible process is antithetical to
unitarity, which tells us first and foremost that when the state of a
system is initially in a pure state then it will remain forever after
pure.  The kind of irreversibility associated with eternal black holes
may be disturbing, but does not violate unitarity in that the state of
the entire system including the black hole can be pure, even though
the state interior to the hole may be inaccessible. This is a
limitation on the {\em evolution} of states rather than a breakdown of
unitarity.  The problem arises for evaporating black holes.  If
Hawking's original calculation is correct than initial pure states
{\em do} evolve into mixed states.
% -- after evaporation there is no inaccessbile system
%left to purify the state.

One potential solution to the problem is that information will
leak out, preserving unitarity, but only when the black-hole reaches 
the Planck scale---the
point at which Hawking's semi-classical calculation breaks down.
Alternatively, the black hole could stop evaporating at the Planck
scale, trapping all the information in a small {\em remnant}.  Both
these solutions were considered unsatisfactory, because it was believed
that the Planck-sized black hole or remnant had to contain all the
original information that went into the hole.  

This is problematic since the well-known black hole entropy formula
of Bekenstein and Hawking \cite{entropyb,entropyh}
\begin{equation}
S_{\rm BH}=4\pi M^2
\end{equation}
(where we work in Planck units $\hbar=c=G=1$) tells us that a small
black hole of mass $M_f$ cannot contain all the entropy of a larger
initial black hole with mass $M \gg M_f$.  But since we are discarding
semi-classical calculations for tiny black holes anyway, this is not terribly
convincing.  
% **** I put this back the old way: Two reasons.  First, the sentence
% was terrible (what ``researchers'' are we talking about?)  Second,
% if we are going to talk about back-action, we need to add a reference.
% That's fine but I don't know what to put.
%But since researchers can discard
%the semi-classical calculations for tiny black holes anyway, as well
%as claim that back-reaction effects will alter the situation even
%for larger black-holes, this is not terribly convincing. 
% {\tt this argument is not really used AFAIK, why include it?}

A stronger objection to a small black hole containing a large amount of 
information, $M^2$ bits, is that the
final burst of radiation in which all this information is released
needs to last a time of order $M^4$.
This is such a long time that one is effectively left
with a stable remnant\cite{acn87,CarlitzWilley}.  Stable remnants are
implausible, because if they contain all the original information of
the black hole, then there are of order $M^2/M_P^2$ different species
of remnants, with $M_P$ the Planck mass, and this huge degeneracy
would have a noticeable impact on low energy physics due to coupling
between remnants and gravitons or soft quanta.

In this letter we attempt to clarify the situation by making a careful
information-theoretic statement about the problem and then showing
that one of the main objections to unitarity is flawed.  Recent
results in quantum information theory\cite{locking,randomization} tell
us that it is not true that simply because information escapes only at
the end of a process that all the information must reside in the small
object remaining at the end of the process.  Instead, we show that the
information can reside in the large Hilbert space of the quanta which
have escaped, but this information is inaccessible.  It is ``locked''
and only becomes available with access to the small number of
remaining quanta, which act in a manner reminiscent of a cryptographic
key.  This is a purely quantum effect and cannot be understood using
only classical information theory.  In the classical case, information
must either reside outside the black hole, or be left inside.  Locking
information classically requires a key as large as the information to
be concealed.  A quantum key can be much smaller.
Thus, one can have a unitary process such that the black hole
evaporates, but leaks little information until the final stages of
evaporation.  The final remaining quanta act as a key, and when they
are finally emitted they restore the full information that was trapped
in the black hole.

While the locking process might appear to be rather ad hoc, we
will further show that it can be made very natural, and can arise generically.
There are however fresh issues which arise when information is
locked in a black hole---namely, we will see that an observer
with special knowledge about the set of initial states of the
system used to create the black hole can get some information out
of the black hole at early times.  

We thus do not claim to have a complete solution to the information
loss problem, which we suspect will require a greater understanding of
quantum gravity.  We merely wish to clarify the black hole paradox in
light of new effects in quantum information theory.  The current
discussion is based on presuppositions originating from classical
reasoning which are simply untrue once one takes into account the
quantum nature of information.

If we believe general relativity, unitarity must break down for
evaporating black holes because Hawking's calculation explicitly tells
us the radiation from a evaporating black hole is thermal, and
therefore independent of the input state.  That is to say, no
information can escape from inside the horizon, even when a black hole
is undergoing evaporation and losing mass.  This is a fundamental
consequence of the disconnected space-time structure of the black
hole.  That Hawking's calculation results in no information escape is
no surprise; the classical causal structure is treated as the
background.

We now formulate in a precise way what it means to
%preserve causality.
have no information escape.  Usually this is expressed
by saying that for all initial pure states $\psi$, ${\cal
S}_t(|\psi\rangle\!\langle\psi |)=\rho$, where $\rho$ does not depend
on $\psi$ and ${\cal S}_t$ is the evolution operator acting on an
external observer's state up until time $t$ (production of Hawking
radiation in this case).  Since we will argue that this is {\em not}
true for $t\rightarrow \infty$ (the overall evolution is unitary) we
consider a time $t$ where the hole has evaporated for a while, but is
still large.

In order to make this more rigorous and physical, and to allow for
small effects due to quantum gravity corrections to Hawking's
semi-classical calculation, let us make the condition for no escape of
information more precise.  Usually one imagines that a single known
state has formed the black hole (an encyclopedia for example).  But
a single state contains no information---information is about 
correlations---it is information {\it about something} and is thus defined over
ensembles.  We should instead imagine a two party game, where one
party A (Alice) forms a black hole from a set of states
$\{|\psi_i\rangle\}$.
The other party B (Bob), observes all the Hawking quanta until time $t$ and
based on measuring the quanta (collectively) tries to guess which
state from the set $\{i\}$ formed the black hole.  As we shall see,
low entropy of emitted Hawking radiation for a given initial state
need not mean information leakage in and of itself.

We want to say that no matter the initial state, the output of the
black hole at $t$ contains nearly no information about the initial
state.  
{\em I.e.}~for all of Bob's measurements $M$ taking 
${\cal S}_t(|\psi_i\rangle\!\langle\psi_i |)$ to classical outcome $j$
\begin{equation}
I(i:j) < \epsilon
\label{eq:smalli}
\end{equation}
where $I(a:b)=H(a)+H(b)-H(ab)$ is the classical mutual information and
$H$ is the Shannon entropy function.  $I$ quantifies the amount of
information which leaks out of the black hole, and equation
(\ref{eq:smalli}) says that there will be little correlation between
the initial states and Bob's guess of what these initial states were.
We can write this in a mixed classical-quantum notation as
\begin{equation}
I_c\left(i:{\cal S}_t(\frac{1}{d}\sum_{i=1}^d|\psi_i\rangle\!\langle\psi_i |)\right) < \epsilon \ .
\label{quantumform}
\end{equation}
$I_c(i:\rho)$ is defined as the maximum classical mutual information
about $i$ that can be obtained by measuring $\rho$.

In classical information theory the following always holds:
\begin{equation}
I(xy:z) - I(x:z) \le I(y:z) \le H(y)
\label{classicalrelation}
\end{equation}
In other words, the
additional information about $z$ gained by having both $x$ and $y$
instead of only having $x$ is no bigger than the entropy contained in
$y$.
%\footnote{The reason the information gain can be {\em less} than
%the entropy in $y$ is that $y$ could contain information redundant to that
%in $x$}.  
%Viewed one way, it could be said that this relation is why
%mutual information is defined the way it is.  
It is this classical relation
that gives us the  false intuition that if nearly no information has escaped
a black hole up until time $t$ then the remaining small hole must
contain nearly all the information.  Quantum mechanically this
intuition is simply wrong.

Following \cite{locking} we define the states
\begin{equation}
\rho= \frac{1}{dn}\sum_{i=1}^d\sum_{j=1}^n 
U_j|i\rangle\!\langle i|U_j^\dag
\end{equation}
\begin{equation}
\rho'= \frac{1}{dn}\sum_{i=1}^d\sum_{j=1}^n 
U_j|i\rangle\!\langle i|U_j^\dag\otimes |j\rangle\!\langle j|
\end{equation}
with $|i\rangle$ and $|j\rangle$ forming orthonormal sets, and the $U_j$'s
are a set of $n$ different unitary operators acting on a $d$-dimensional
Hilbert space.  The difference between these is that the second state
includes a classical label (encoded in the orthonormal set of $|j\rangle$'s)
telling which of the $n$ possible unitaries $U_j$ was applied.
Comparing the difference in accessible classical information when one does 
or does not have $j$ it has been shown \cite{locking,randomization} 
that for certain choices of $n,d$ and the $U_j$'s
\begin{equation}
I_c(i:\rho')-I_c(i:\rho) \gg \log{n}\ .
\end{equation}
That is to say that quantumly (\ref{classicalrelation}) can be violated
by an arbitrarily large amount.  The information of ``which $i$'' is locked
by not having access to the $j$.  Here, the number of different $j$'s can 
be small.  In particular, by choosing $n=(\log d)^3+k$ 
and the $U_j$'s at random (over the Haar measure), we have for large $d$ 
and constant $C$ \cite{footnoteimprovement}
\begin{equation}
I_c(i:\rho') = \log{d}
\label{logd}
\end{equation}
\begin{equation}
I_c(i:\rho) < \delta = C^{-k}
\label{delta}
\end{equation}

Eq. (\ref{delta}) is of the form (\ref{quantumform}) if only we
assume the action of the black hole ${\cal S}_t$ is to perform one of a set
of random unitary operators and apply this evolution to an ensemble of orthogonal states $|i\rangle$.

Now let us rewrite $\rho'$ as
\begin{equation}
\rho_{B\!H}= \frac{1}{dn}\sum_{i=1}^d\sum_{j=1}^n 
(U_j|i\rangle\!\langle i|U_j^\dag)_B \otimes |j\rangle\!\langle j|_H\ .
\label{rhobh}
\end{equation}
This is simply an assignment of Hilbert spaces.  
If ${\cal H}_H$ is considered to be inside a black hole and inaccessible 
to $B$, his remaining state would be ${\rm Tr}_H(\rho_{B\!H})=\rho$.  Thus, 
the small $n$-dimensional
Hilbert space of the black hole has kept the mutual information $B$
has about $i$ low.  If the black hole completely evaporates, 
Hilbert space ${\cal H}_H$ is transferred to $B$ and his
final state becomes $\rho'$, with accessible information
$\log{d}$---all the information is restored.  Or the information can be locked forever by a remnant which hides
the value of $j$ living in ${\cal H}_H$.

The evolution taking $\rho_0=(1/d)\sum_i |i\rangle\!\langle i|$ 
to $\rho_{B\!H}$ 
%(\ref{rhobh}) 
is not unitary.  The
entropy in the sum over $j$'s has appeared out of nowhere.  We
would like to find an evolution with the same $S_t$ describing the state
outside the black hole, while not producing extra entropy.
We replace $\rho_{B\!H}$ with 
\begin{equation}
\rho_{B\!H}'= \frac{1}{dn}\sum_{i=1}^d\sum_{j=1}^n \sum_{k=1}^n 
(U_j|i\rangle\!\langle i|U_k^\dag)_B \otimes |j\rangle\!\langle k|_H\ .
\label{purified}
\end{equation}
Since ${\rm Tr}_H \rho_{B\!H}' = {\rm Tr}_H \rho_{B\!H}$
the evolution outside the black hole is the same, and $I_c$ is
unchanged.  The difference is that the state still inside the black hole 
is now entangled with the external state rather than classically
correlated with it.  After complete evaporation $B$ has a
superposition of $j$'s instead of a mixture.  He can measure in the
$|j\rangle$ basis and collapse the superposition yielding $\log{d}$
information as before.
Thus, we now have a completely unitary process by which a small black
hole can lock the large amount of information that it originally contained.
When the small hole finally finishes evaporating all the information is regained.

There is one serious problem with the above analysis.  It depends
crucially on the black hole having been formed from an ensemble of
orthogonal states chosen with uniform probability and spanning nearly
the entire Hilbert space of the hole.  The ``locking'' phenomenon is
extremely vulnerable to {\em coding}.
%\footnote{This can be seen from
%the fact that the classical capacity is large}. 
In other words, if the ensemble is
restricted to spanning a space of slightly less than $d/n$ dimensions,
then nearly all the information is accessible (due to the packing lemma \cite{igorpacking}).  
The actual amount is about $\log{d}-\log{n}$ or the same as we'd expect
classically from (\ref{classicalrelation}).  Thus, if Alice creates
the black hole not with all possible states, but instead restricts the set 
of states she uses, and Bob knows the restricted set of states,
then Bob will be able to guess the value of $i$ before the black hole
has fully evaporated.  Equivalently, putting many copies of the same state $i$ into many black holes
in order to repeat the two-party game will allow Bob to distinguish the value of $i$ because
effectively, the total Hilbert space is being restricted to one with identical copies of the
same state.

There is an elegant way of looking at the black hole information problem
based on arguments by Susskind \cite{susskind-thorlacius}.  
If evolution takes a state outside of the
light cone (from A to B for example), and our theory is relativistically 
invariant, then there
exists a reference frame in which the state has evolved from an initial
copy at A, to two copies of the state, one at A and one at B.
Such an evolution cannot be unitary---it violates the no-cloning
theorem \cite{nocloning}.  In the case of the black hole, one finds
a space-like hypersurface which is well-away from the singularity,
yet intersects almost all the outgoing Hawking radiation as well as the
infalling matter which formed the black hole inside the apparent horizon.
This hypersurface contains two copies of the state.
Thus if information eventually escapes the black hole,
the no-cloning theorem (and hence, unitarity and linearity), would be violated.
We thus have the amusing situation that if no information escapes from
the black hole, unitarity is violated, yet if information escapes from a black
hole, unitarity once again appears to be violated.

In light of information locking, we see that this argument is not
strictly true.  In our model, the full state cannot be reconstructed
from the outgoing radiation until the final burst of radiation (and
this burst of radiation is not captured by any hypersurface which
avoids the singularity).  In other words, one can have information
eventually leak out at the Planck scale in such a way that the black
hole cannot be used as a universal cloning machine.  However, due to
the coding argument above, one can use the black hole to clone some
subspace of the full Hilbert space---still disastrous for quantum
theory, but we hope this clarification may lead to advances removing
even this smaller violation of causality.

Turning back to our model, we should point out that the actual state
of the outgoing radiation and internal states are not of the form
(\ref{purified}), but rather, are quantum fields in a thermal state
outside the hole, correlated with the quantum states of the black hole
(geometry).  Our $S_t$ should be taken only as a simplified model,
designed to clarify our central argument.  For any initial state, one
can find a realistic mapping which takes the output state to one for
which any small number of quanta appear thermal, thus reproducing the
semi-classical Hawking result at any instant.  It should be
possible to find a mapping where this is true for any initial
state, however, the totality of the emitted radiation cannot appear
thermal, containing phase correlations over many quanta and having
small total entropy. This is necessarily in a unitary theory and can
be attributed to the semi-classical analysis not
taking into account quantum effects of the geometry, {\em e.g.}\ back-reaction
effects.  By some estimates, the semi-classical calculation completely
breaks down once the black hole has evaporated a fraction of order
$M^{-2/3}$ of its mass \cite{page-unitary-evap}. Note that at
intermediate times the entropy of the radiation for any particular
initial state $|i\rangle$ will be low, although it is high enough that
observing the radiation does not allow one to distinguish which state
$i$ was used to form the black hole.  This low entropy is sometimes
taken to mean that information has leaked out. However, as we have
seen, this is only true when the number of potential input states is
restricted.

It is also true that the remaining small black hole that ``locks'' the
information until quantum effects dominate needs to contain
information growing as $3 \log\log{d}$.  Additionally, applying the
analysis of Carlitz and Willey\cite{CarlitzWilley} one finds that the
lifetime of the remnant will not be of order $M^4$ as before, but
rather, much shorter, of order $(\log M)^2$.  Since these quantities
can grow without bound with the size of the initial black hole, the
original complaint about the small hole needing to hold a large amount
of information still vexes.  Fortunately, logarithmic growth is
extremely slow.  For a solar mass black hole holding at most $10^{38}$
bits of information, the evaporation process can be semi-classical
until the black hole is tiny, of size around $20$ Planck masses.
There may also be a cosmological limit on just how large a black hole one
really needs to worry about. 
Taking Lloyd's estimate \cite{lloyd} of
$10^{90}-10^{120}$ bits as the total bits in the universe, then all
this information could be locked by a black hole with a mass of only
$40$ Planck masses.  One can easily imagine quantum effects coming into
play for a hole that small.

Finally, let us turn to potential mechanisms which might generate such
states.  It might appear that the state we have used is highly
artificial, but this is not the case.  Even unitaries chosen at
random will work, thus one expects rather generic
mechanisms to yield such states.  All one needs is that the knowledge of 
which unitary acted remain inside the black hole.  Of course, we are
not claiming a precise mechanism for this which would presumably
requires a fuller understanding of a quantum theory of gravity.  Our
purpose is to refute standard presuppositions about black hole
information and to suggest a possible form of evolution resulting from
a potential theory of quantum gravity.  We have shown that there is an
evolution $S_t$ that leads to a state $\rho_{B\!H}$ where $B$ has
nearly no information about what state formed the black hole, but if $B$ has 
access to the small system $H$ then $B$ has complete information.  If the
evolution of a black hole were $S_t$ then both unitarity and causality
could be preserved without requiring a small nearly-evaporated black
hole to be able to hold all the information that ever fell into its
large ancestor.  Information can still leak out of the black hole
if the initial set of states is known to be restricted. 
This is related
to the fact that the purely quantum information measure the{\em coherent
information} \cite{SN1996}cannot be locked (which can be shown using
standard entropy inequalities). Clearly further work needs to be done
clarifying precisely which measure of information should be used when
analyzing black holes and causality, a detail largely ignored in the
literature.  We hope that this letter may point future attempts at
reconciling black hole information loss in useful directions.

The authors would like to thank K. Horodecki, R. Oliveira,
J. Preskill, L. Susskind and A. Winter for helpful
discussions.  Thanks to the Newton Institute for support and providing
an environment of scientific interchange.  JAS thanks ARO contract
DAAD19-01-C-0056.  JO thanks EU grants PROSECCO and COSLAB.

\end{document}